\documentclass{article}

\usepackage{arxiv}
\usepackage[utf8]{inputenc} % allow utf-8 input
\usepackage[T1]{fontenc}    % use 8-bit T1 fonts
\usepackage{url}            % simple URL typesetting
\usepackage{booktabs}       % professional-quality tables
\usepackage{nicefrac}       % compact symbols for 1/2, etc.
\usepackage{microtype}      % microtypography
\usepackage[hidelinks]{hyperref}
\hypersetup{
  colorlinks = true, %Colours links instead of ugly boxes
  urlcolor = black, %Colour for external hyperlinks
  linkcolor = black, %Colour of internal links
  citecolor = black %Colour of citations
}

\usepackage{doi}

\usepackage{graphicx}
\usepackage{subcaption}
\usepackage{caption}
\usepackage{multirow}
\usepackage{amsthm} %theorem
\usepackage{amsmath}
\usepackage{amsfonts}
\usepackage{longtable}
\usepackage{array}
\usepackage{makecell}
\usepackage{placeins}
\usepackage[ruled,vlined]{algorithm2e}

\graphicspath{{Images/}}
\usepackage{xcolor}

\newtheorem{definition}{Definition}

\newcolumntype{M}[1]{>{\centering\arraybackslash}m{#1}}

\title{OutCenTR: A Novel Semi-supervised Framework for Predicting Exploits of Vulnerabilities in High Dimensional Datasets}

\author{Hadi Eskandari*,
        Michael Bewong*,
        Sabih ur Rehman*\\ % <-this % stops a space\\
        {*School of Computing, Mathematics and Engineering, Charles Sturt University, Australia}}

\begin{document}
\maketitle

\begin{abstract}
An ever-growing number of vulnerabilities are reported every day. Yet these vulnerabilities are not all the same; Some are more targeted than others. Correctly estimating the likelihood of a vulnerability being exploited is a critical task for system administrators. This aids the system administrators in prioritizing and patching the right vulnerabilities. Our work makes use of outlier detection techniques to predict vulnerabilities that are likely to be exploited in highly imbalanced and high-dimensional datasets such as the National Vulnerability Database. We propose a dimensionality reduction technique, OutCenTR, that enhances the baseline outlier detection models. We further demonstrate the effectiveness and efficiency of OutCenTR empirically with 4 benchmark and 12 synthetic datasets. The results of our experiments show on average a 5-fold improvement of F1 score in comparison with state-of-the-art dimensionality reduction techniques such as PCA and GRP.
\end{abstract}


\section{Introduction}
Estimating the likelihood of a vulnerability being exploited is a critical challenge in computer security and requires well thought-out solutions. A vulnerability is a weakness in a system which often occurs as a result of poor code design, inevitable bugs, or underlying hardware limitations. These weaknesses allow threat actors to perform actions that can cause system failure and loss of data or privacy. For example, in 2021 a vulnerability named Log4Shell was found in the commonly used logging library, Log4j. Thus, any device and software that used this library were exposed to remote-code execution (RCE) attacks. According to the National Institute of Standards and Technology (NIST)\footnote{https://www.nist.gov}, the number of vulnerabilities discovered is increasing exponentially. For instance, 8,051 vulnerabilities were published in the National Vulnerability Database (NVD) \footnote{https://nvd.nist.gov} in the first quarter of 2022. This is a 25\% increase compared to the same period last year. 

\begin{figure}[ht]
	\centering
	\includegraphics[width=0.95\linewidth]{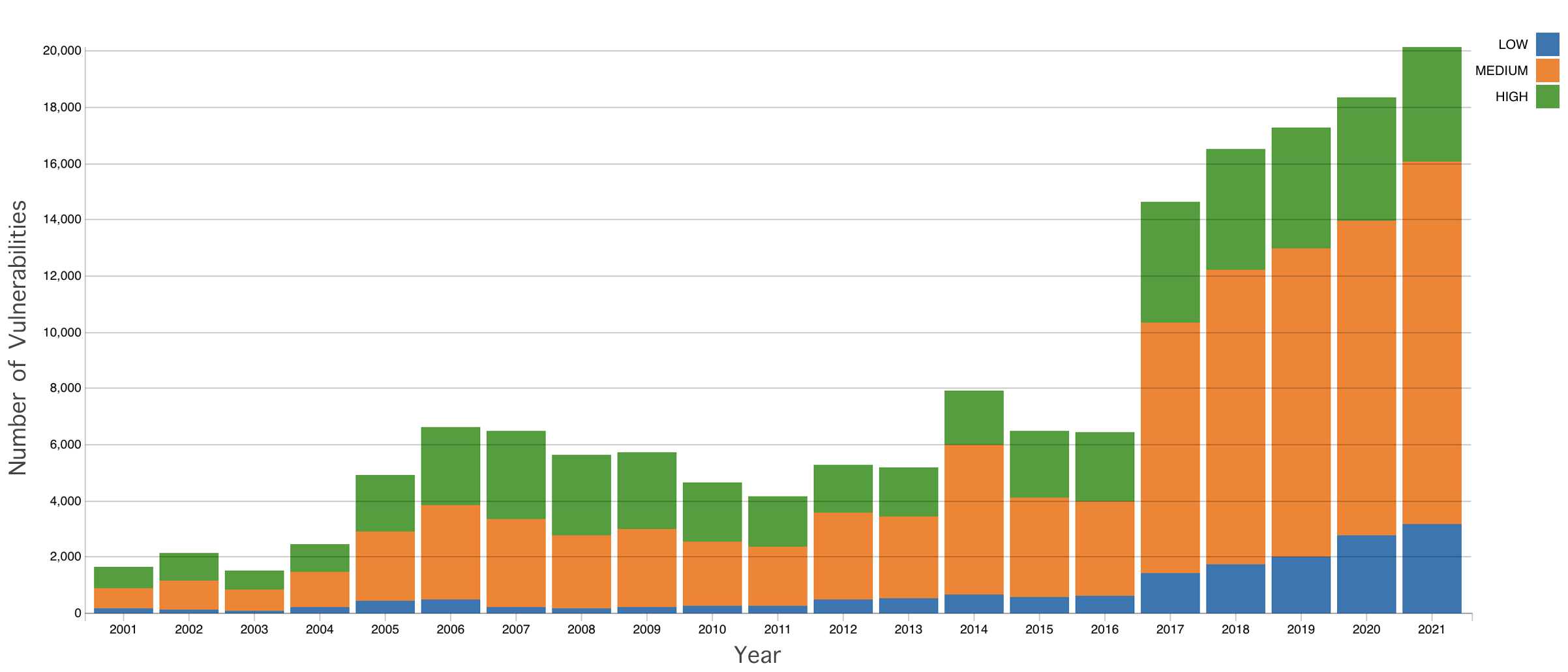}
	\caption{Trend of vulnerabilities}
	\label{fig:CVSS_Distribution}
\end{figure}

It is important to note, however, that not all vulnerabilities are the same. Most vulnerabilities are hardly ever exploited, with only a few leading to attacks \cite{bullough_2017}. A key challenge for system administrators is how to decide which vulnerabilities should be prioritized for effective system hardening, under the constraint of limited resources. The increased rate of discovered vulnerabilities and the change in their severity over time can be seen in Figure.~\ref{fig:CVSS_Distribution}. Many studies have attempted to predict software vulnerability exploits using machine learning models. Other Studies \cite{bozorgi_2010,edkrantz_2015,bullough_2017,almukaynizi_2017,huang_2020,bhatt_2021} have used supervised machine learning algorithms such as Support Vector Machines (SVM), Random Forest and K-nearest neighbors among others. However, there are three main limitations with the existing approaches: Firstly, the existing studies make the inherent assumption that labeled datasets for training of these models can easily be curated or obtained. Clearly, this assumption does not hold in real-world scenarios. For example, the NVD database does not label vulnerabilities as exploited or not, instead, linkages between the NVD database and other exploits databases have to be made to infer such labels. The problem is further exacerbated by the exponential growth in vulnerabilities, making it cost-ineffective and futile for the data to be constantly updated and labeled. As a result, catalogs such as Open-Source Vulnerability Database (OSVDB)\footnote{http://www.osvdb.org} which provided accurate and up-to-date information about vulnerabilities shut down due to high cost of maintenance \cite{ruohonen_2017}.

Secondly, most vulnerability databases have high dimensionality which can often impair the model learning process. For example, vulnerability datasets' dimensionality is often in the hundreds as shown in the empirical section, or even in the thousands as demonstrated in the seminal work of \cite{bozorgi_2010}.  At the same time, as demonstrated in the empirical section well-known feature reduction techniques such as Principal Component Analysis (PCA) \cite{Jolliffe_2011} and Gaussian Random Projection (GRP) \cite{bingham_2001} are not suited to addressing this specific problem in the context of exploits of vulnerabilities.

Thirdly, while the development of deep learning techniques such as \cite{huang_2019} have shown great promise, they do not have the inherent interpretability that is needed to support decision-making. In addition, such techniques often require a large amount of labeled data and computing resources making them inapplicable in resource-constrained environments.

In this work, we propose a semi-supervised machine learning approach to overcome the above-mentioned challenges. In particular, we propose an outlier-centric feature reduction framework (OutCenTR) that leverages outlier detection as a means of identifying vulnerabilities that are likely to be exploited. The intuition of this approach is based on the fact that outlier detection techniques have been used successfully in detecting anomalous data points in \cite{kollios_2003}; detect fraudulent events in finance \cite{ahmed_2016}; disease detection \cite{wong_2003} \cite{lin_2005}; and in computer security for intrusion detection \cite{wenlishang_2015}. However, the application of outlier detection to predict the exploit of vulnerabilities is neither trivial nor straightforward. For example, existing outlier detection techniques are ineffective in high dimensional datasets due to the loss of differentiability of records in a high dimensional space~\cite{beyer_1997}. To this end, we propose OutCenTR which is not only effective for predicting exploit of vulnerabilities but also demonstrates its effectiveness in general-purpose outlier detection. We make the following contributions:

\begin{itemize}
    \item Propose a novel outlier detection framework as a means of predicting exploits of vulnerabilities. To the best of our knowledge, this is the first work of its kind to adopt a semi-supervised machine learning approach to solving this problem.

    \item Design a centroid-based feature reduction technique, OutCenTR, to address the challenge of high dimensionality in datasets. Furthermore, we show the effectiveness and efficiency of OutCenTR on four diverse real-world high-dimensional datasets and twelve synthetic datasets.
    
    \item Compare the performance of OutCenTR to the state-of-the-art dimensionality reduction techniques such as PCA and GRP and demonstrate the improvements of our approach.
\end{itemize}

The remainder of our paper is structured as follows: Section \ref{Sec:Relatedwork} reviews the state-of-the-art and other related methods. Section \ref{Sec:Preliminaries} provides problem formulation and theoretical insights. Section \ref{Sec:OutCenTR} proposes our solution framework. Section \ref{Sec:Results} provides a summary of the results and findings of our experimentation, and in Section \ref{Sec:Conclusion} we conclude our work.

\section{Related work \label{Sec:Relatedwork}}

Some studies have explored the use of machine learning models in categorizing the software vulnerability types~\cite{das_2021,huang_2019}. In \cite{das_2021} authors have used deep learning models to classify Common Vulnerabilities and Exposures (CVE) entries into distinct Common Weakness Enumeration (CWE) types to quickly help identify the type of the weakness. They tested their model for two distinct periods to account for concept drift. Similarly in \cite{huang_2019} authors have used Term Frequency/Inverse Document Frequency (TF-IDF) to build a model to classify the category of the disclosed vulnerability based on textual information of the NVD dataset. Their deep learning model was built with 900 input layers and 5000 hidden layers and the number of features extracted from the NVD dataset was ($m=1024$).

Works such as~\cite{zhang_2011,sapienza_2017,yin_2022,chen_2019} have focused on estimating the time to vulnerabilities or exploit of vulnerabilities. In \cite{zhang_2011} researchers developed vendor-specific models with support vector machines (SVMs) using data from the NVD database to predict the time to the next vulnerability for major vendors separately. Another proposed model in \cite{chen_2019} predicts if and when vulnerability will be exploited. This framework updates its database through new CVE and NVD databases and using Twitter conversations, in order to visualize the information in a graph. In \cite{sapienza_2017} authors have created a framework to alert imminent threats using social media data sources such as Twitter and the Darknet websites and white-hat hacker communities. Further, authors in \cite{yin_2022} have designed a sliding window learning model to predict the time to exploit using the CVSS version 2.0 from the NVD dataset and linked it to the ExploitDB database.

Closely related to our work, are the works in~\cite{bozorgi_2010,edkrantz_2015,sabottke_2015,jacobs_2019,yin_2020} which focus on estimating the likelihood of an exploit of a vulnerability. In \cite{bozorgi_2010} authors have used a support-vector machine (SVM) model to predict the possibility of an exploit being developed for a known vulnerability. 

As shown in \cite{tang_2019} there is a multivariate dependence structure amongst different vulnerability risks and their respective exploits. In \cite{edkrantz_2015} authors have focused on predicting exploits being developed for newly found vulnerabilities, based on the previous exploit patterns. Whereas in \cite{sabottke_2015} authors have used information from social media such as Twitter and the textual data available on the NVD or CVE records to enrich their model created for response prioritization based on the likelihood of the exploit being developed. In \cite{huang_2020} authors have trained a Random Forest model to calculate the exploit tendency of disclosed vulnerabilities using data from 1999 to 2019. In \cite{jacobs_2019}, the authors proposed a logistic regression model built using CVE and NVD databases to overcome the shortcomings of the CVSS scoring system. Furthermore, in \cite{yin_2020} authors propose a BERT neural network model trained on Wikipedia to detect the predictability of exploits using NVD dataset vulnerability description.

It is evident from the above-mentioned approaches, that it is usually assumed that the labeled dataset is often available, however, this may not be a reasonable assumption in real-world scenarios \cite{pang_2021,meng_2016,hanif_2021}. For example, it can take several months for an exploited vulnerability to be assessed and registered as an exploit. Until such a label has been issued, the vulnerability is assumed unexploited. This can lead to poor inaccurate models. Further, existing works often use artificially balanced datasets. That is, they often evade the reality of the imbalanced nature of vulnerability datasets by accumulating large amounts of data over a long period and then generating a balanced dataset from this to train models. This creates unnaturally optimistic models which may not be applicable to future data. While deep-learning models generally cope well with imbalanced datasets, these models lack interpretability, which many security domain experts consider a major risk \cite{pang_2021,lin_2020}. Our proposed framework uses a semi-supervised approach that does not require a large body of labeled datasets, is well suited for imbalanced datasets, and benefits from the interpretability associated with outlier detection models. In the following section, we present preliminaries and formalize the distance-based feature selection problem. Readers are encouraged to refer to Table \ref{Tab:Used_Symbols} for symbols used to define the proposed framework in the following sections. 

%%%%%%% Symbols %%%%%%%%%
\begin{table}[t]
\centering
\small
\caption{Symbols and Notations}
\label{Tab:Used_Symbols}
\begin{tabular}{|c|l|}
\hline
\textbf{ Symbol }      & \textbf{ Description }\\ \hline
\Xhline{2\arrayrulewidth}
$R$          & A dataset \\ \hline
$r$          & Records of the dataset $R$ \\ \hline
$R'$         & Transformed dataset $R$  \\ \hline
${\bf r}$    & Vector representation of $r$ \\ \hline
$\tilde{R}$  & The mass of the dataset $R$  \\ \hline
$m$          & dimensions of the dataset \\ \hline
$\Phi(R)$    & An outlier detection algorithm  \\ \hline
$n$          & Number of records \\ \hline
$A$          & A set of dataset attributes  \\ \hline
${\bf C}$    & The centroid vector\\ \hline
$th$         & The outlier threshold  \\ \hline
$\mathbb{C}$ & Outlier detection context \\ \hline 
$O$          & The outliers of a dataset  \\ \hline
$dist$       & An distant metric function \\ \hline
$\bar{O}$    & The inliers of a dataset  \\ \hline
$\mathbb{S}$ & The distinguishability score\\ \hline
$ARank$      & Attribute rank of $A$ \\\hline
\end{tabular}
\end{table}
%%%%%%%%%%%%%%%%%%%%%%%%%

\section{Distance-Based feature selection for Outlier Detection\label{Sec:Preliminaries}}

Let $R$ be a dataset with $n$ records \emph{i.e.} $R= \{r_1,\cdots,r_n\}$, where each record $r$ is a projection on the set of attribute $A = \{a_1,\cdots,a_m\}$. For simplicity, we assume all attribute values in the dataset $R$ are normalized to $[0,1]$. Also, let $\Phi(R)$ be an outlier detection function that returns the set of outliers in $R$. In outlier detection, the objective is to find the record $r_i$ in the dataset $R$ that deviates from other records $r_j$. The deviation is often measured via a distance metric denoted $dist$. A record is classified as an outlier when its deviation $dist(r_i, \tilde{R})$ exceeds some threshold $th$, where $\Tilde{R}$ is the mass of $R$ e.g. centroid.  In this regard, outlier detection can be contextual and application specific. For example, the amount of network traffic during the sales season e.g. Christmas, new year holidays, or other expected peaks is usually higher than the other periods. Even though traffic is high during these periods, this may not be an outlier due to the expected high traffic during this time. An equal amount of traffic, however, could be deemed as an outlier event if it happens outside of the expected period. We, therefore, introduce the term \emph{context} to capture the scope of the outlier detection task for dataset $R$. A \emph{context} is formally defined as follows:
\\
\begin{definition}[Context $\mathbb{C}$]
\label{Def_Context}
Let $R=\{r_1,\cdots,r_n\}$ be a dataset for a specific outlier detection task \emph{e.g.} intrusion detection, the context of the dataset $R$ captures the scope of the outlier detection and is denoted by the pair $\mathbb{C}(R) = \langle dist, th\rangle$, where $dist$ is the distance function which evaluates the distance of each record $r_i$, and $th$ is the threshold, above which a record $r_i$ is considered an outlier.
\end{definition}

\begin{definition}[Outlier Detection $\Phi$]
\label{Def_OutlierDetection}
Given a set of records $R=\{r_1,\cdots,r_n\}$ and $context$ $\mathbb{C}(R) = \langle dist, th\rangle$, an outlier detection function $\Phi(R)$ returns a set of records \{$r_i$\} such that  $dist(r, \tilde{R})>th$ for all $r \in \{r_i\}$.
\end{definition}

We denote the true set of outliers in $R$ by $O$ and the set of inliers by $\bar{O}$ such that $O\cap \bar{O} = \emptyset$ and  $O, \bar{O} \subset R$ thus $ O + \bar{O} = R$ and $O << \bar{O}$. In this paper, we treat a vulnerability dataset as $R$ and each vulnerability record that may or may not be exploited as $r \in R$. According to \cite{almukaynizi_2017} and corroborated by our experiments, only a small fraction of vulnerabilities are ever exploited \emph{i.e.} $O << \bar{O}$. We posit that vulnerabilities that are exploited possess specific inherent characteristics that make them more amenable to exploit than the vulnerabilities that are not exploited. However, as previously seen, vulnerability datasets often have high dimensionality which makes existing outlier detection techniques ineffective. In this work, we propose a novel feature reduction technique called Outlier Centric Feature Reduction (OutCenTR) which enhances the effectiveness of outlier detection techniques on high-dimensional datasets. Formally, the outlier-centric feature reduction problem is defined as follows:
\\
\begin{definition}[Problem Formulation]
\label{Def_ProblemFormulation}
Let $R$ be a dataset where each $r \in R$ is a projection on the set of attributes $\{a_1,\cdots,a_m\}$, and $\mathbb{C}(R) = \langle dist, th\rangle$ denote the context. The challenge is to transform $R$ into $R'$ where $r'\in R'$ is a projection on the set of attributes $\{a_k\}$ and $\{a_k\} \subseteq \{a_1,\cdots,a_m\}$ such that the number of true outliers detected from the transformed dataset $\Phi(R')$ is greater than the number of true outliers detected from the original dataset $\Phi(R)$ \emph{i.e.} $\{\Phi(R')\cap O\} \geq \{\Phi(R)\cap O\}$. 
\end{definition}

\section{Outlier-Centric Feature Reduction\label{Sec:OutCenTR}}
In this section, we propose our model called Outlier-Centric feaTure Reduction (OutCenTR) which differs from other dimensionality reduction techniques such as GRP and PCA. In GRP, the algorithm is based on Johnson-Lindenstrauss lemma \cite{johnson_1984} where a dataset with dimension $m$ is projected to a subspace with $n$ dimensions where $n << m$ using a random matrix with $k \times m$ size \cite{bingham_2001}. In PCA, there is a transformation of $m$ number of  correlated variables into $n$ uncorrelated linear projections of the original variables where $n << m$ \cite{wold_1987}. Both GRP and PCA are not effective in outlier detection though useful for classification and exploratory data analysis tasks \cite{ahn_2019}. 

\begin{figure}[t]
    \centering
    \includegraphics[width=.8\linewidth]{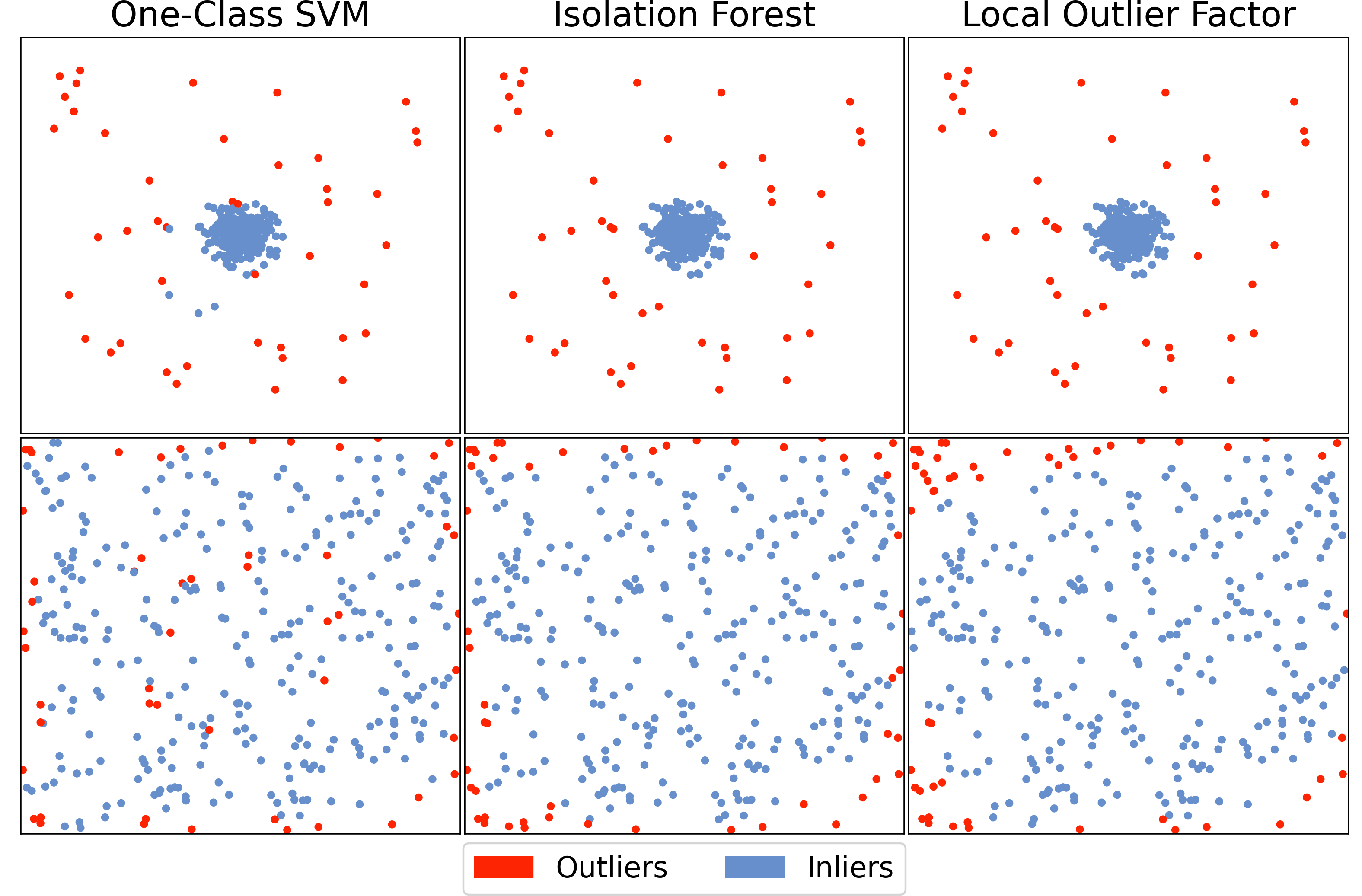}\hfill
    \caption{Outlier detection in high vs low dimensions}
    \label{fig:OutlierDetection}
\end{figure}

We make the observation that when attributes are carefully selected based on their distinguishability of outliers and inliers, we can achieve better outlier detection results. For example, Figure.~\ref{fig:OutlierDetection} shows that when the features of the dataset are carefully chosen in the first row, the outliers are easier to separate, but when they are not carefully selected in the second row, outlier detection can be much harder. In high-dimensional datasets, the presence of dimensions, such as the second row in Figure.~\ref{fig:OutlierDetection} adds noise to the modeling task. Our main goal in OutCenTR is to identify which subspaces within the dataset maximize the difference between outliers and inliers.

OutCenTR utilizes the average representation of a sample of outliers and inliers respectively to estimate the most distinguishing features of the data. Following Definition \ref{Def_Context}, we use the centroid \emph{i.e.} the mean of the dataset $R$ to represent the mass of the dataset $\tilde R$ as is commonly used in statistical studies \cite{barnett_1978}. However, a key observation made is that when the centroid is calculated using both outliers and inliers, the true mass of the dataset is often degraded. Based on this observation, we leverage a few labeled samples to calculate the centroid of inliers and outliers separately. From these two centroids, we are then able to determine the attributes that contribute most to the characterization of the outliers by comparing the respective pairs of attribute values in both centroids. The following formal definitions are needed for OutCenTR.
\\
\begin{definition}[Dataset Mass $\tilde{R}$]
Given a dataset $R$, its mass $\tilde{R}$ is given by its centroid {\bf C} as follows: 
\begin{equation}
    \label{f:DatasetMass}
    \tilde{R}:= {\bf C} = {\frac {1}{n}}\sum _{r \in R} ^ n  {\bf{r}},
\end{equation}

where {\bf r} is the vector representation of {r} and {\bf C} is the vector representation of the centroid.
\end{definition}

\begin{figure*}[t]
\centering
  \begin{subfigure}{\linewidth}
    \centering
    \includegraphics[width=.8\linewidth]{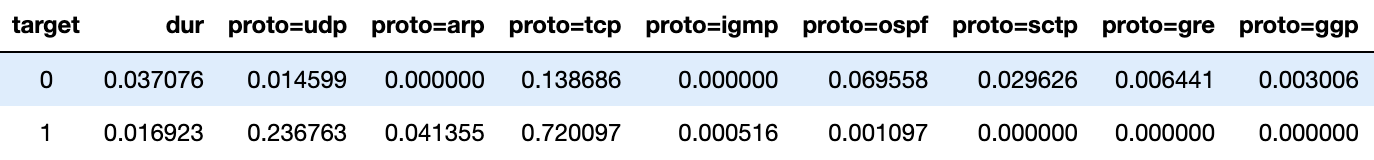}
    \caption{\label{fig:Centroid-Calculation}}
  \end{subfigure}\par\medskip
  \begin{subfigure}{\linewidth}
    \includegraphics[width=1\linewidth]{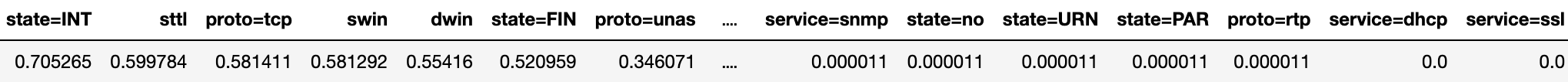}
    \caption{\label{fig:Score-Calculation}}
  \end{subfigure}\par\medskip
  \caption{Calculation of Centroid and Attribute Score}
  \label{fig:IntrusionDataSet}
\end{figure*}

We denote the centroid of the set of outliers and inliers by ${\bf C}_{O}$ and ${\bf C}_{\bar O}$ respectively. Figure~\ref{fig:Centroid-Calculation} illustrates the result of a centroid calculation for an intrusion detection dataset. In the figure, each row corresponds to the centroids where `target' denotes the label \emph{i.e.} $0$ represents the centroid for inliers, ${\bf C}_{\bar O}$, and $1$ represents the centroid for outliers, ${\bf C}_{O}$. 

To define how features in a dataset $R$ are ranked based on their ability to distinguish outliers from inliers, we present the distinguishability score $\mathbb{S}$.
\\
\begin{definition}[Distinguishability Score $\mathbb{S}$]
Given a dataset $R$ and the set of attributes $\{a_1,\cdots,a_m\}$, the distinguishability score $\mathbb{S}_j$ for an attribute $a_j$ is calculated as follows:
\begin{equation}
    \label{f:Score}
    \mathbb{S}_j = |\mathbf{C}_{O}.a_j - \mathbf{C}_{\bar{O}}.a_j|,
\end{equation}

where ${C}_{x}.a_j$ is the corresponding value of attribute $a_j$ in the centroid ${C}_{x}$.
\end{definition}

The distinguishability scores for all attributes can be ordered to produce a ranking defined as the attribute rank, $ARank$, as follows.
\\
\begin{definition}[Attribute Rank $ARank$]\label{Def-ARank}
Given a dataset $R$ and the set of attributes $\{a_1,\cdots,a_m\}$ and their corresponding distinguishability scores $\{\mathbb{S}_1,\cdots,\mathbb{S}_m\}$, the attribute rank $ARank(R)$ of the dataset $R$ is the ordered set $\{\mathbb{S}_h\preceq,\cdots,\preceq\mathbb{S}_r \}$ such that the distinguishability score $\mathbb{S}_h \geq \mathbb{S}_r $. 
\end{definition}

Following the definition above, we denote the top $t$ attributes based on $ARank$ as $ARank_t$, where $t$ is a user-defined parameter. For the purpose of attribute ranking and selection, the score is defined as:

\begin{definition}[Attribute Score $score.a_j$]
Given the attributes of the dataset $R$, the score $score.a_j$ for the attribute $a_j$ is the average distance of that attribute from the centroid:
\begin{equation}
\label{f:AttributeScore}
score.a_j = {\frac {1}{n}}\sum _{i = 1 } ^ n { r_i . a_j - centroid_p . a_j},
\end{equation}
where $p \in \{O, \bar O \}$.
\end{definition}

Figure.~\ref{fig:Centroid-Calculation} shows the calculation of the centroid where attributes that have a greater mean distance from the centroid such as `state=INT' and `sttl' will have higher ${score.aj}$ than others. Figure.~\ref{fig:Score-Calculation} shows the resulting attribute rank of the same dataset. Algorithm~\ref{Tab:Algorithm} summarises the steps in OutCenTR where the feature reduction phases are accomplished in Steps 1 - 5 and the outlier detection phase is accomplished in Step 6.

%%%%%%Algorithm Table%%%%%%%%
\begin{algorithm}[t]
\tiny
\caption{OutCenTR}
\label{Tab:Algorithm}
\SetAlgoLined
\textbf{Input}: dataset ${R}$\\
\textbf{Processes}: Outlier detection with OutCenTR\\
\textbf{Output}: A set of outliers ${O}$\\
\textbf{Step 1:} Normalize the dataset $R$ to $[0,1]$ \\
\textbf{Step 2:} Compute the mass of the dataset $\Tilde{R}$ i.e. centroid (equation~\ref{f:DatasetMass})\\
\textbf{Step 3:} Using the context $\mathbb{C}$, calculate the distinguishability score $\mathbb{S}$ (equation~\ref{f:Score}) of each attribute ${a_j}$\\
\textbf{Step 4:} Create ${ARank}$ as an ordered set of pairs of $<a_j, \mathbb{S}>$\\
\textbf{Step 5:} Calculate scores for all attributes according to equation~\ref{f:AttributeScore} \\
\textbf{Step 6:} Select the configured top $x$ attributes from the $ARank$\\
\textbf{Step 7:} Transform the dataset $R$ into $R'$ by selecting the configured top $x$ attributes based on the $ARank$\\
\textbf{Step 8:} Use the $\Phi(R')$ to return the set of found outliers {$O$} in the dataset
\end{algorithm}
%%%%%%Algorithm call%%%%%%%%

\section{Empirical evaluation\label{Sec:Results}}

Our experiments were run on a 3 GHz 10-Core Intel Xeon machine with 64GB memory. To evaluate our framework four benchmark datasets and twelve synthetic datasets were used. The benchmark datasets consist of a vulnerability dataset \emph{NVD}, \emph{Network Intrusion} dataset, \emph{Fraud Detection} dataset and \emph{Census} dataset\footnote{\emph{Network Intrusion}, \emph{Fraud Detection} and \emph{Census} are obtained from \cite{guansong_2021}, while \emph{NVD} is curated by us from the NVD website and ExploitDB databases at exploit-db.com}. The synthetic datasets were generated using the datasets package of ScikitLearn library \footnote{http://www.scikit-learn.org}. These packages were used to generate a random dataset with specified separability, number of attributes, and records. Table \ref{Tab:DataSetCharacteristics} shows characteristics of our datasets.

The categorical data were encoded as numeric values. All numerical values were scaled to be between 0 and 1. Each dataset is split into  training (80\%) and testing (20\%) sets. Each dataset had a ground truth label which was used in the calculation of evaluation scores.

%%%%%%Characteristics of the datasets%%%%%%%%

%The original table is above, in case needed
\begin{table}[ht]
\caption{Dataset characteristics}
\label{Tab:DataSetCharacteristics}
\small
\centering
\begin{tabular}{|l|l|l|l|}
\hline
\multicolumn{1}{|c}{\textbf{Dataset}} & \multicolumn{1}{|c|}{\textbf{\# Records}} & \multicolumn{1}{c|}{\textbf{\# Attributes}} & \multicolumn{1}{c|}{\textbf{In vs. Outlier}} \\ \Xhline{2\arrayrulewidth}
NVD               & 61501        & 286            & 95\%, 5\%  \\ \hline
Network Intrusion & 95329        & 197            & 98\%, 2\%  \\ \hline
Fraud detection   & 284807       & 30             & 99\%, 1\%  \\ \hline
Census            & 299285       & 501            & 94\%, 6\%  \\ \hline
Synthetic         & 1000 or 2000 & 50, 100 or 200 & 95\%, 5\%  or 90\%, 10\%\\ \hline
% \rule{0pt}{8pt}SYN-2             & 1000      & 50           & 95\%, 5\%  \\ \hline
% \rule{0pt}{8pt}SYN-3             & 2000      & 50           & 90\%, 10\% \\ \hline
% \rule{0pt}{8pt}SYN-4             & 1000      & 50           & 90\%, 10\% \\ \hline
% \rule{0pt}{8pt}SYN-5             & 2000      & 200          & 95\%, 5\%  \\ \hline
% \rule{0pt}{8pt}SYN-6             & 1000      & 200          & 95\%, 5\%  \\ \hline
% \rule{0pt}{8pt}SYN-7             & 2000      & 200          & 90\%, 10\% \\ \hline
% \rule{0pt}{8pt}SYN-8             & 1000      & 200          & 90\%, 10\% \\ \hline
% \rule{0pt}{8pt}SYN-9             & 2000      & 100          & 95\%, 5\%  \\ \hline
% \rule{0pt}{8pt}SYN-10            & 1000      & 100          & 95\%, 5\%  \\ \hline
% \rule{0pt}{8pt}SYN-11            & 2000      & 100          & 90\%, 10\% \\ \hline
% \rule{0pt}{8pt}SYN-12            & 1000      & 100          & 90\%, 10\% \\ \hline
\end{tabular}
\end{table}

%%%%%%%%%%%%%%%%%%%%%%%%%%%%%%%%%%%%%%%%%%%%%

One-Class SVM (OCSVM), Isolation Forest (iForest), and Local Outlier Factor (LOF) were chosen for the purpose of this experiment. OCSVM was configured to use the non-linear kernel (RBF) as the best-performing kernel. The ratio of outliers in the dataset was used for the `nu' parameter. The `gamma' parameter was entered as auto. For iForest, the contamination parameter was set to the ratio of the outliers in the dataset, and the `max samples' parameter was left as auto. For LOF, the contamination parameter was set to the ratio of the outliers in the dataset. To indicate outlier detection, the `novelty' parameter was set to false. The `algorithm' parameter was left on auto. We compared two main settings: (a) the baseline outlier detection models, and (b) detection through OutCenTR\footnote{The source code for OutCenTR proposed in this paper will be available on \href{https://github.com/HEskandari/outcentr-research}{GitHub} once the final version of the paper is published.} In all the experiments involving OutCenTR, the default number of selected attributes \emph{i.e.} parameter $t$ in $ARank_t$ was $10\%$ (\emph{c.f.} Definition~\ref{Def-ARank}).

\begin{figure*}[t]
\centering
  \begin{subfigure}{.475\textwidth}
    \centering
    \includegraphics[width=.6\textwidth]{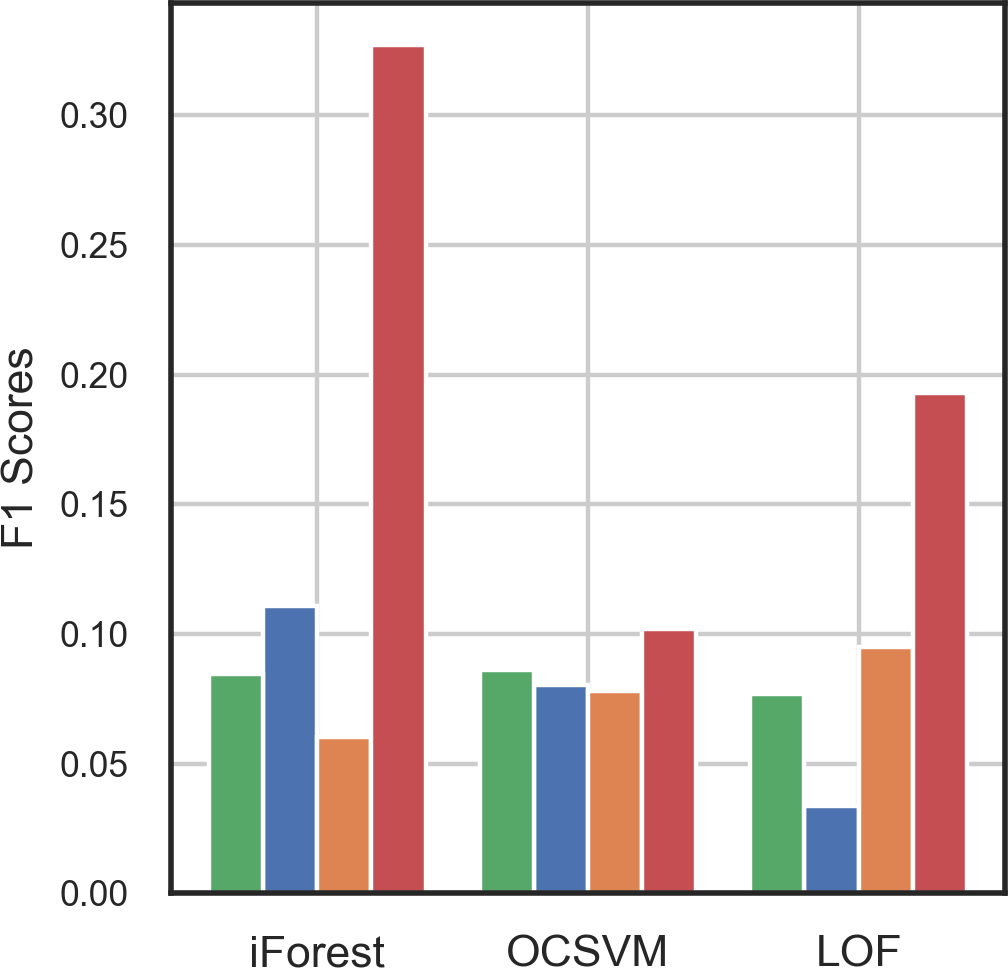}
    \caption{\label{fig:Metric-F1}}
  \end{subfigure}
  \hfill
  \begin{subfigure}{.475\textwidth}
    \centering
    \includegraphics[width=.6\textwidth]{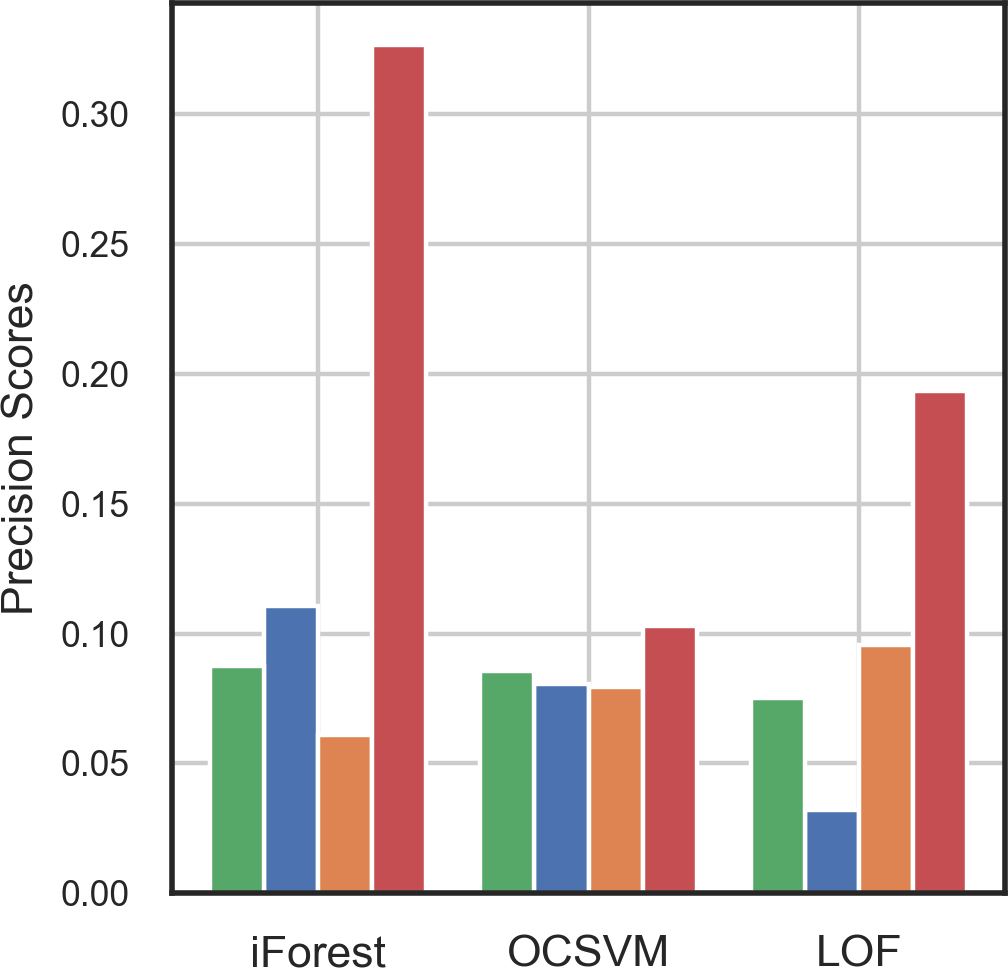}
    \caption{\label{fig:Metric-Precision}}
  \end{subfigure}
  \vskip\baselineskip
  
  \begin{subfigure}{.475\textwidth}
    \centering
    \includegraphics[width=.6\textwidth]{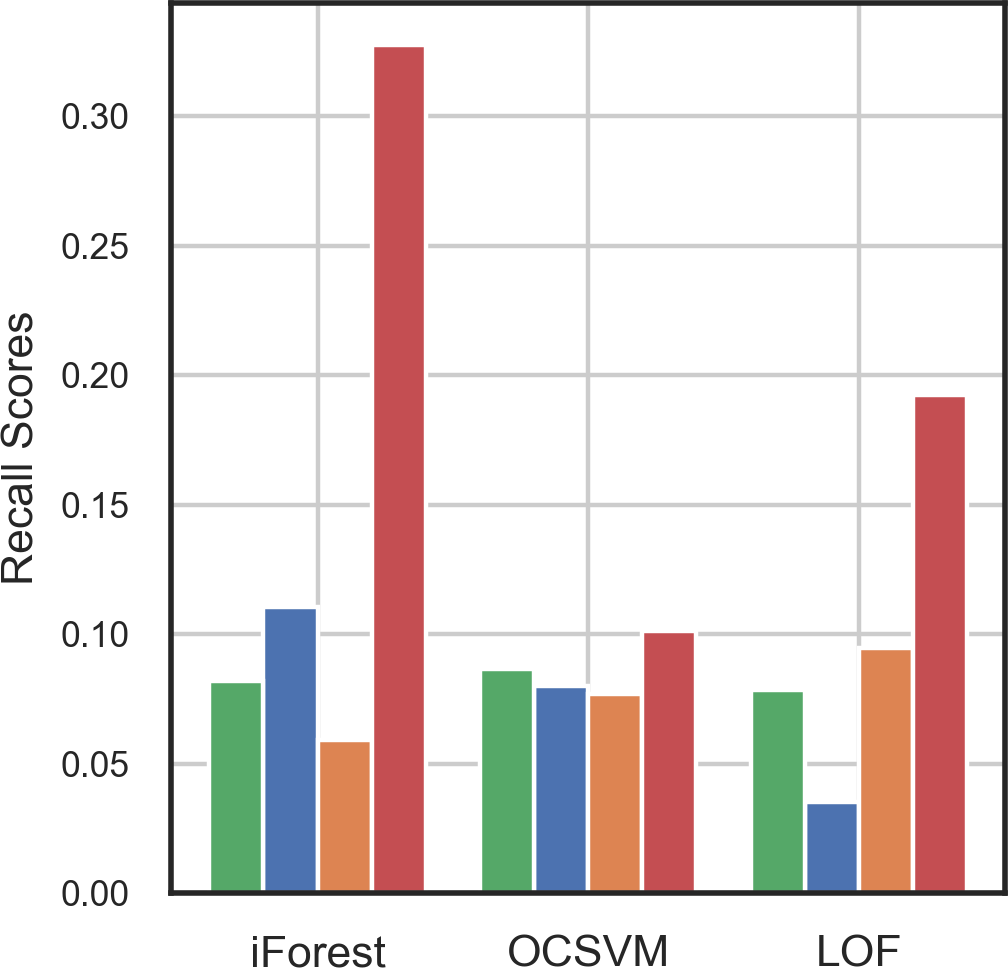}
    \caption{\label{fig:Metric-Recall}}
  \end{subfigure}
  \hfill
  \begin{subfigure}{.475\textwidth}
    \centering
    \includegraphics[width=.6\textwidth]{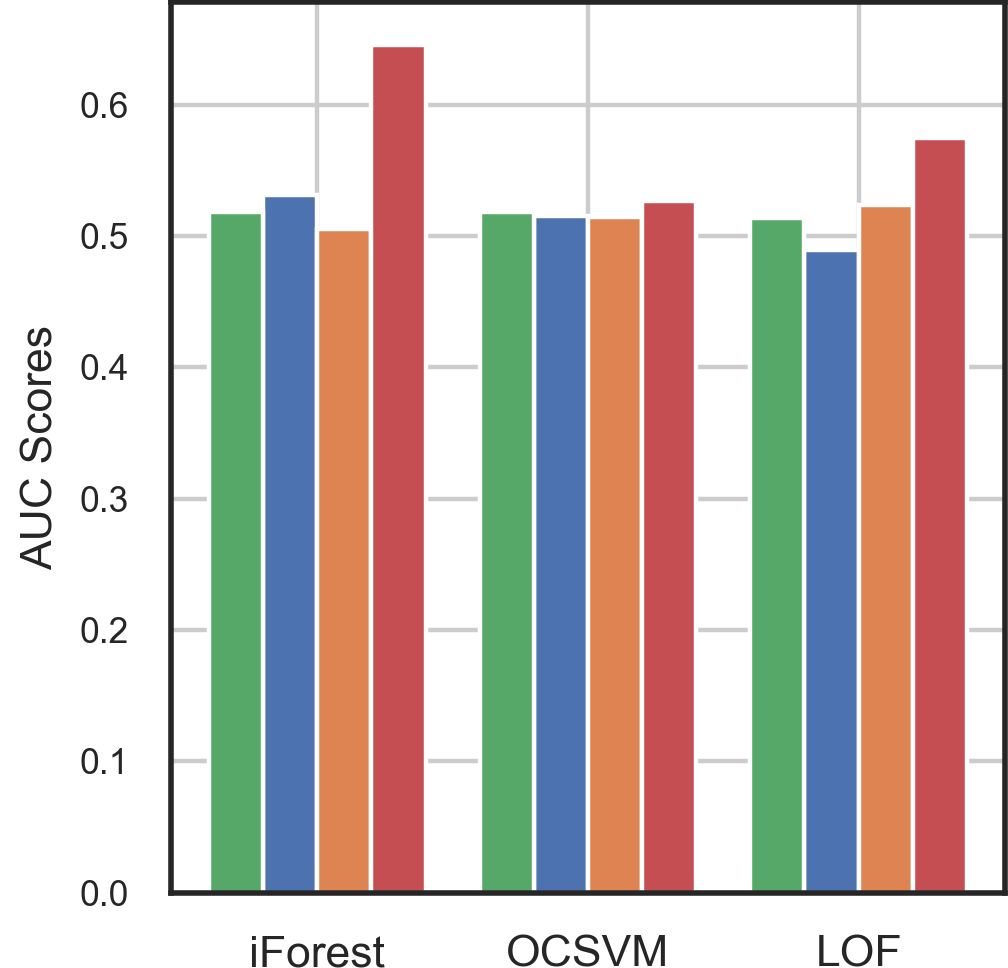}
    \caption{\label{fig:Metric-AUC}}
  \end{subfigure}
  \hfill
  \par\medskip
  \begin{subfigure}{.4\textwidth}
    \centering
    \includegraphics[width=\textwidth]{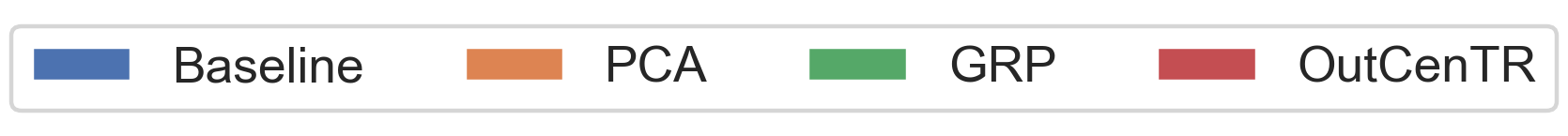}
  \end{subfigure}
  \caption{Comparison of OutCenTR with PCA and GRP}
  \label{fig:NVDMetricsComparison}
\end{figure*}

\subsection{Effectiveness in the prediction of exploit of vulnerability}
To assess the effectiveness of OutCenTR in selecting relevant features in comparison with PCA and GRP within the context of vulnerability detection, the NVD dataset was used. An equal number of features were used for OutCenTR, PCA, and GRP ($m=28$) to reduce the feature space and create a training dataset. The resulting trained models were used to predict outliers of the NVD test dataset using iForest, OCSVM, and LOF. For comparison, the F1 score (Figure.~\ref{fig:Metric-F1}), Precision (Figure.~\ref{fig:Metric-Precision}), Recall (Figure.~\ref{fig:Metric-Recall}), AUC score (Figure.~\ref{fig:Metric-AUC}) metrics were gathered. The chart shows the performance of the selected models in four different configurations: the baseline (in blue) \emph{i.e.} the original dataset vs. transformed dataset through PCA (orange), GRP (green), and OutCenTR (red). The highest F1 score, precision, recall, and AUC score through all three outlier detection models were achieved using OutCenTR. The highest F1 score of OutCenTR is $\approx5$ times higher than PCA and GRP. Whilst OutCenTR in Figure.~\ref{fig:NVDMetricsComparison} shows significant improvements over other techniques, it is tempting to view the overall scores as modest. However, it is important to note that the objective of this experiment is to demonstrate the performance of OutCenTR in comparison with conventional dimensionality reduction techniques. Further, this work is the first of its kind to demonstrate a recall and precision of $\approx35\%$ each using a semi-supervised framework for real-time prediction of exploits.

\subsection{Effectiveness in outlier detection \label{Sec:EffectivenessComparison}}

To determine the effectiveness of OutCenTR for general-purpose outlier detection tasks, we performed further experiments using \emph{NVD}, \emph{Network Intrusion}, \emph{Fraud Detection}, and \emph{Census} datasets. The experiments focus on the effectiveness using F1 score, precision, and recall. Table \ref{Tab:ResultsTable} shows the result. The optimum F1 score, precision, and recall in all scenarios appear in bold. The optimum F1 score of iForest and OCSVM were achieved by OutCenTR. To understand how various parameters such as dataset dimension, number of records, and the ratio of outliers vs inliers affect the performance of OutCenTR, a similar experiment was run on a set of synthetically generated datasets: OutCenTR generally achieves a much better precision and recall average with iForest and OCSVM models: overall, the precision and recall of OutCenTR has improved by $\approx 3$ times and $\approx2$ times respectively. In both of our experiments, LOF generally showed poor performance. As stated in \cite{cheng_2019} LOF is sensible to local outliers and it may yield better results in ensemble settings and when combined with a global outlier detection model. Due to space limitations, the result table for synthetic datasets is omitted from this paper.

% \usepackage{multirow}
% \newcolumntype{M}[1]{>{\centering\arraybackslash}m{#1}}

\begin{table}[t]
\tiny
\centering
\caption{Experimental results with benchmark datasets}
\label{Tab:ResultsTable}

\begin{tabular}{|c|l|M{9mm}|M{9mm}|M{9mm}|}
\hline
{\textbf{Dataset}} & {\textbf{Model}} & {\textbf{F1}} & {\textbf{Recall}} & {\textbf{{Precision}}} \\ \hline
\Xhline{1\arrayrulewidth}
\multirow{6}{*}{\rule{0pt}{5pt}{NVD}} &
    \rule{0pt}{5pt}Isolation Forest & 6.87\% & 4.98\% & 11.11\% \\ \cline{2-5} &
    \rule{0pt}{5pt}OCSVM & 4.09\% & 3.05\% & 6.21\% \\ \cline{2-5} &
    \rule{0pt}{5pt}LOF & 1.89\% & 1.44\% & 2.75\% \\ \cline{2-5} &
    \rule{0pt}{5pt}Isolation Forest (OutCenTR) & \textbf{28.01\%} & \textbf{20.71\%} & \textbf{43.29\%} \\ \cline{2-5} &
    \rule{0pt}{5pt}OCSVM (OutCenTR) & \textbf{8.86\%} & \textbf{6.42\%} & \textbf{14.29\%} \\ \cline{2-5} & 
    \rule{0pt}{5pt}LOF (OutCenTR) & \textbf{3.70\%} & \textbf{2.73\%} & \textbf{5.74\%} \\ \hline
\Xhline{1\arrayrulewidth}

\multirow{6}{*}{\rule{0pt}{5pt}{Network Intrusion}} &
    \rule{0pt}{5pt}Isolation Forest & 1.46\% & 1.58\% & 1.36\% \\ \cline{2-5} &
    \rule{0pt}{5pt}OCSVM & 5.20\% & 5.18\% & 5.22\% \\ \cline{2-5} &
    \rule{0pt}{5pt}LOF & \textbf{35.52\%} & \textbf{34.68\%} & \textbf{36.41\%} \\ \cline{2-5} &
    \rule{0pt}{5pt}Isolation Forest (OutCenTR) & \textbf{26.65\%} & \textbf{26.35\%} & \textbf{26.96\%} \\ \cline{2-5} &
    \rule{0pt}{5pt}OCSVM (OutCenTR) & \textbf{8.19\%} & \textbf{8.33\%} & \textbf{8.06\%} \\ \cline{2-5} &
    \rule{0pt}{5pt}LOF (OutCenTR) & 8.24\% & 8.78\% & 7.75\% \\ \hline
\Xhline{1\arrayrulewidth}

\multirow{6}{*}{\rule{0pt}{5pt}{Fraud Detection}} &
    \rule{0pt}{5pt}Isolation Forest & 12.92\% & 82.47\% & 7.01\% \\ \cline{2-5} &
    \rule{0pt}{5pt}OCSVM & 12.89\% & 80.41\% & 7.01\% \\ \cline{2-5} &
    \rule{0pt}{5pt}LOF & 0.00\% & 0.00\% & 0.00\% \\ \cline{2-5} & 
    \rule{0pt}{5pt}Isolation Forest (OutCenTR) & \textbf{13.98\%} & \textbf{90.72\%} & \textbf{7.57\%} \\ \cline{2-5} &
    \rule{0pt}{5pt}OCSVM (OutCenTR) & \textbf{13.97\%} & \textbf{87.63\%} & \textbf{7.59\%} \\ \cline{2-5} &
    \rule{0pt}{5pt}LOF (OutCenTR) & \textbf{0.31\%} & \textbf{2.06\%} & \textbf{0.17\%} \\ \hline
\Xhline{1\arrayrulewidth}

\multirow{6}{*}{\rule{0pt}{5pt}{Census}} &
    \rule{0pt}{5pt}Isolation Forest & 4.93\% & 4.87\% & 4.99\% \\ \cline{2-5} &
    \rule{0pt}{5pt}OCSVM & 5.71\% & 5.67\% & 5.75\% \\ \cline{2-5} &
    \rule{0pt}{5pt}LOF & 1.97\% & 1.93\% & 2.02\% \\ \cline{2-5} &
    \rule{0pt}{5pt}Isolation Forest (OutCenTR) & \textbf{8.30\%} & \textbf{8.15\%} & \textbf{8.45\%} \\ \cline{2-5} &
    \rule{0pt}{5pt}OCSVM (OutCenTR) & \textbf{17.21\%} & \textbf{17.00\%} & \textbf{17.44\%} \\ \cline{2-5} &
    \rule{0pt}{5pt}LOF (OutCenTR) & \textbf{9.38\%} & \textbf{9.27\%} & \textbf{9.50\%} \\ \hline
\end{tabular}
\end{table}

\subsection{Evaluation of Feature Importance}

\begin{figure*}[t]
\centering
  \begin{subfigure}{.475\textwidth}
    \centering
    \includegraphics[width=1\textwidth]{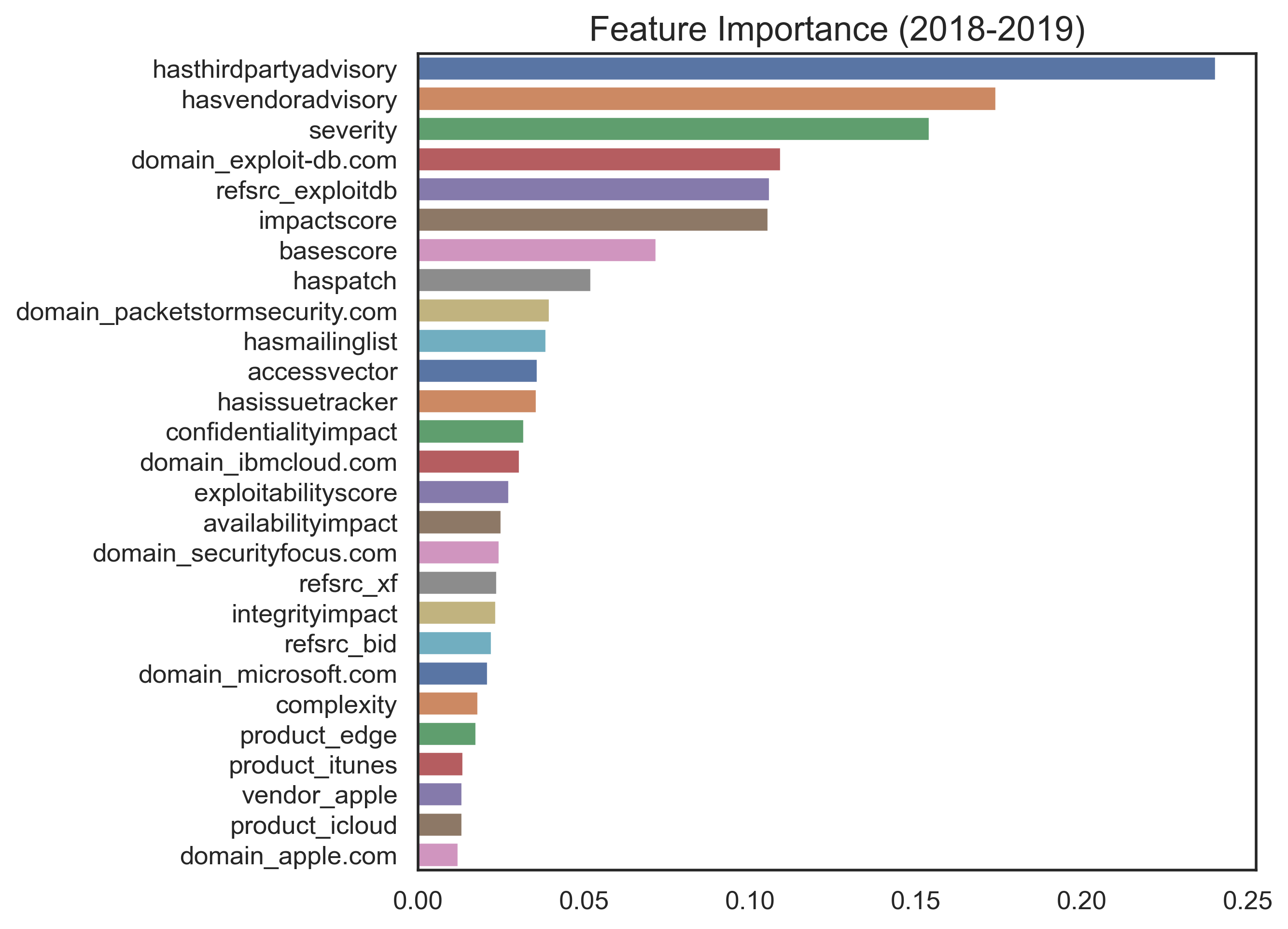}
    \caption{\label{fig:Feature-Importance-2018-2019}}
  \end{subfigure}
  \hfill
  \begin{subfigure}{.475\textwidth}
    \centering
    \includegraphics[width=1\textwidth]{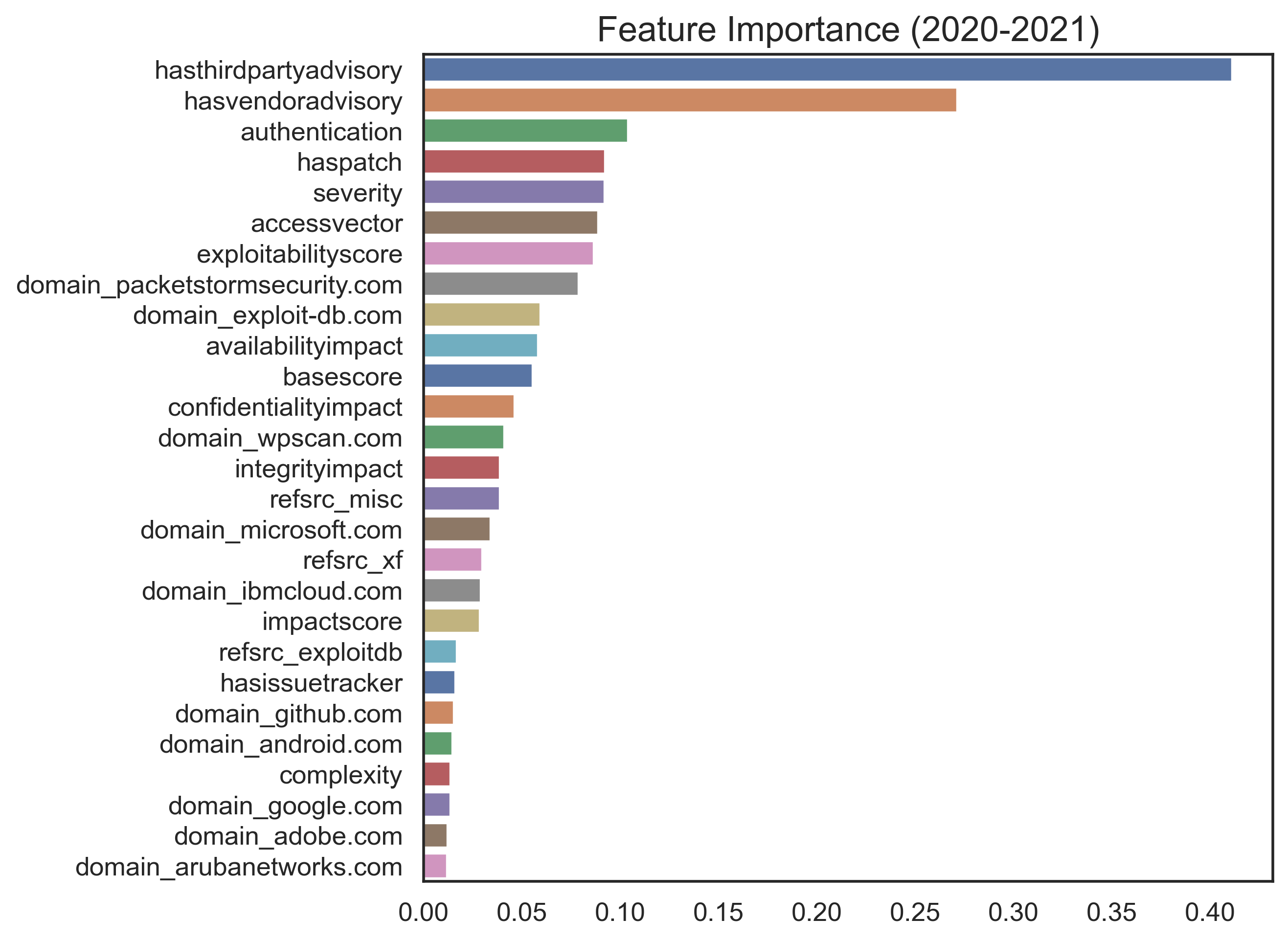}
    \caption{\label{fig:Feature-Importance-2020-2021}}
  \end{subfigure}
  \caption{Attribute Importance comparison}
\end{figure*}

In this experiment, we aim to demonstrate the effect of concept drift and how OutCenTR adapts to changing patterns in a dataset. We ran OutCenTR on the NVD dataset from two separate date ranges: one for the period between 2018 - 2019 (Figure.~\ref{fig:Feature-Importance-2018-2019}) and another test for date range of 2020-2021 (Figure.~\ref{fig:Feature-Importance-2020-2021}). We observed that the selected attributes and their degree of impact for the purpose of vulnerability detection change over time. For example, the `Severity' which was the third impacting attribute in the Figure.~\ref{fig:Feature-Importance-2018-2019} is moved to the fifth position and has a lower rank. Similarly, the `Authentication' attribute has a higher rank in Figure.~\ref{fig:Feature-Importance-2020-2021} which was not deemed as an impacting attribute at all based on the data from the previous range. We believe the OutCenTR feature ranking can be used to determine the best attributes when designing a machine learning model.

\begin{figure*}[ht]
\centering
  \begin{subfigure}{.475\textwidth}
    \centering
    \includegraphics[width=.7\textwidth]{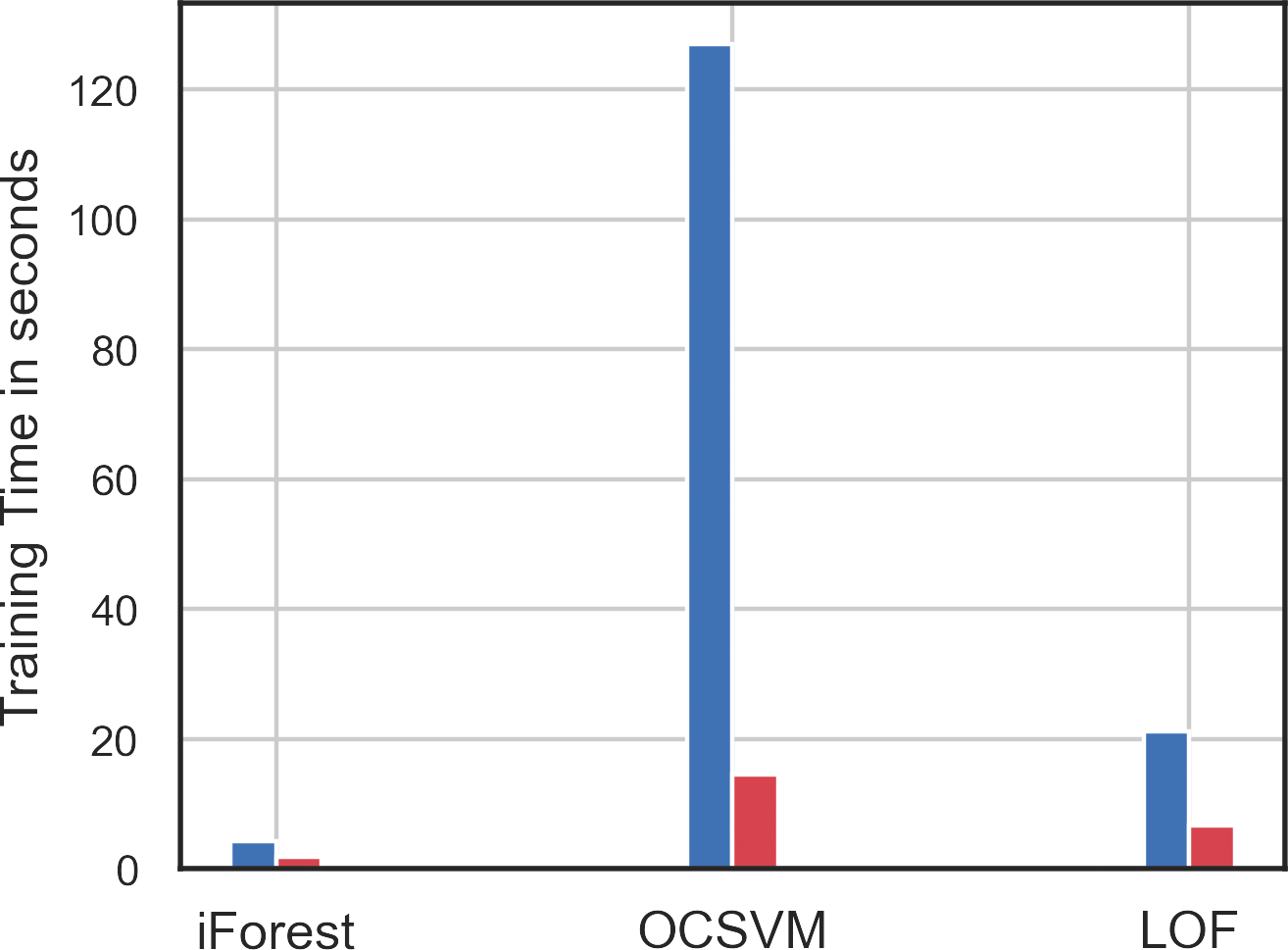}
    \caption{\label{fig:TimeComparison-Fit}}
  \end{subfigure}
  \hfill
  \begin{subfigure}{.475\textwidth}
    \centering
    \includegraphics[width=.7\textwidth]{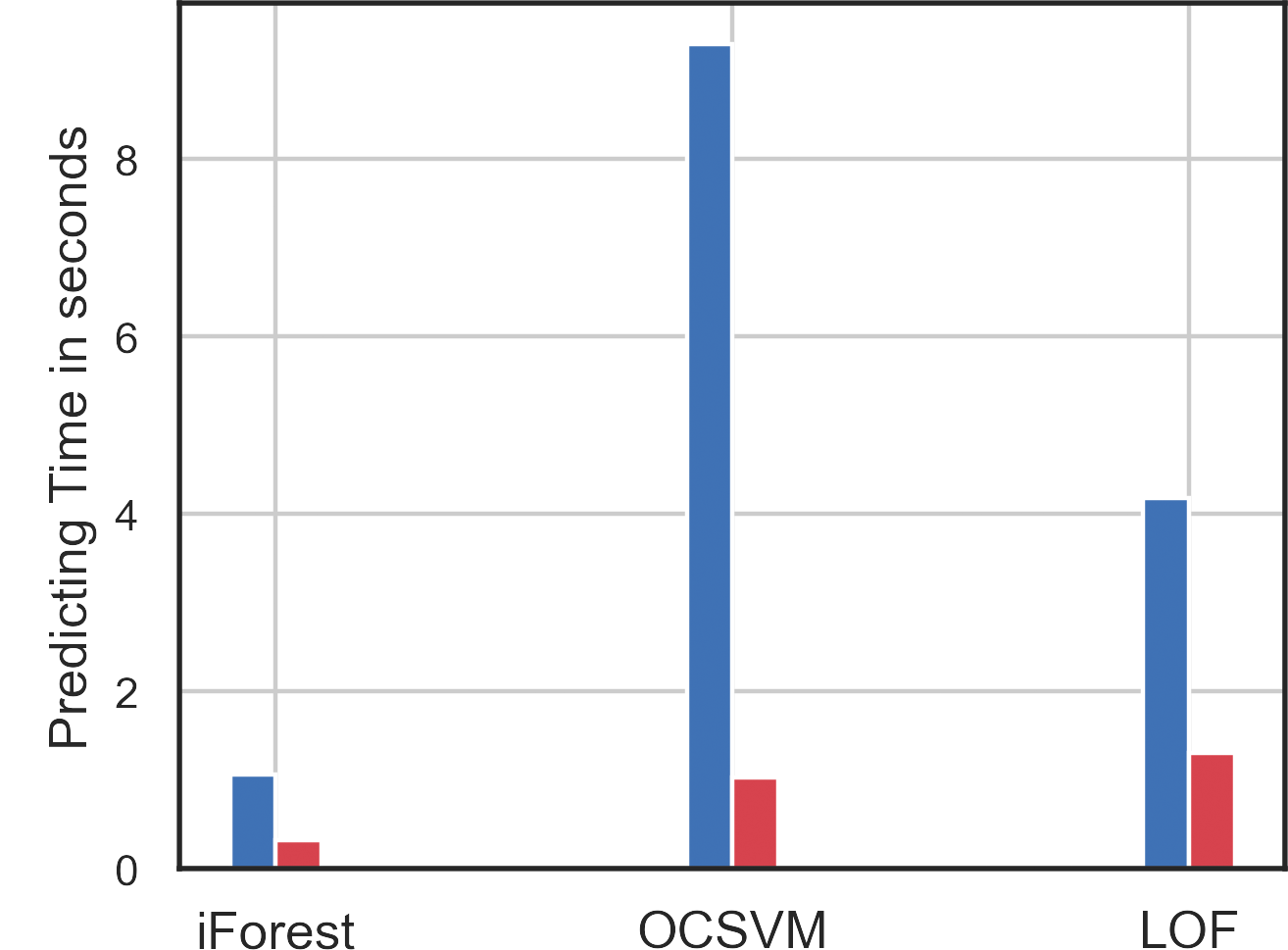}
    \caption{\label{fig:TimeComparison-Predict}}
  \end{subfigure}
  \hfill
  \par\medskip
  \begin{subfigure}{.475\textwidth}
    \centering
    \includegraphics[width=.5\textwidth]{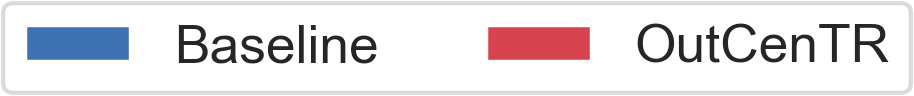}
  \end{subfigure}
  \caption{Baseline time comparisons}
  \label{fig:TimeComparison}
\end{figure*}

\subsection{Time-efficiency comparison\label{Sec:TimeEfficiencyExpr}}

To measure the time efficiencies gained through feature reduction by OutCenTR, we measured the algorithms' training and predicting times in our experiments. The result of this experiment can be seen in Figure.~\ref{fig:TimeComparison}. The experiment was done in two settings: the models trained and tested on the baseline dataset (blue) with no feature reductions applied, and through OutCenTR (red). For both training time and predicting time, OutCenTR performs better than the baseline. For the NVD dataset and the OCSVM, the difference is dramatic: training and prediction ran $\approx7$\ times and $\approx9$ times faster respectively. Due to the algorithmic complexity of OCSVM, very high dimensional datasets such as Census ($m=501$) make OCSVM extremely slow. We noted that the training of the baseline model for the census dataset took more than two hours (8,292 seconds) but through OutCenTR the training time was reduced to about half an hour (2,273 seconds).

\section{Conclusion \label{Sec:Conclusion}}

In this study, we proposed a framework for outlier detection algorithms to detect exploits of vulnerabilities. We address the curse of dimensionality by proposing a new dimensionality reduction technique, OutCenTR, and empirically evaluate its effectiveness and efficiency on four real benchmark and twelve synthetic datasets. When compared to the mainstream dimensionality reduction techniques like PCA and GRP we observed a 5-fold improvement in the F1 score. As part of future work, we are exploring the possibility of using other techniques in the calculation of the mass of the dataset and its effect on determining the important predictors.

\clearpage

\bibliographystyle{IEEEtran}
\bibliography{References} 

\end{document}